\documentclass[letterpaper,12pt]{article}
\usepackage{authblk}
\usepackage[utf8]{inputenc}
\usepackage[english]{babel}
\usepackage{amsmath}
\usepackage{amssymb}
\usepackage{graphicx}
\usepackage{multirow}
\usepackage{epsfig}
\usepackage{rotating}
\usepackage{longtable} 
\usepackage{caption}
\usepackage{subcaption}
\usepackage{url}
\usepackage{pdflscape}
\usepackage{lscape}
\usepackage{cancel}
\usepackage{setspace}
\pdfoutput=1
\usepackage{csquotes}
\usepackage{placeins}
\usepackage[left=2cm,top=3cm,right=2cm,bottom=3cm]{geometry}
\usepackage{tabularx}
\usepackage{adjustbox}
\usepackage{color,xcolor}

\usepackage{hyperref}

\usepackage{mathtools}
\usepackage{latexsym}

\title{Two Texture Zeros for Dirac Neutrinos in a Diagonal charged lepton basis}

\author[a]{Yessica Lenis}
\author[b]{John D. Gómez}
\author[a]{William A. Ponce}
\author[b]{Richard H. Benavides}

\affil[a]{Instituto de Física, Universidad de Antioquia, A.A. 1226, Medellín, Colombia.}

\affil[b]{Instituto Tecnol\'{o}gico Metropolitano, Facultad de Ciencias Exactas y Aplicadas, Calle 73 N° 76-354 via el volador, Medell\'{i}n, Colombia.}

\begin{document}
  
\maketitle

\abstract{A systematic study of the neutrino mass matrix $M_\nu$ with two texture zeros in a basis where the charged leptons are diagonal,  and under the assumption that neutrinos are Dirac particles, is carried through in detail. Our study is done without any approximation, first analytically and then numerically.  Current neutrino oscillation data are used in our analysis.  Phenomenological implications of $M_\nu$ on the lepton CP violation and neutrino mass spectrum are explored.}

\section{Introduction}
Although the gauge boson sector of the Standard Model (SM) with the $SU(3)_c\otimes SU(2)_L\otimes U(1)_Y$ local symmetry has been very successful~\cite{dms,PhysRevLett.13.508,GLASHOW1961579,PhysRevLett.19.1264}, its Yukawa sector is still poorly understood. Questions related to this sector such as the total number of families in nature, the hierarchy of the charged fermion mass spectrum, the smallness of neutrino masses, the quark mixing angles, the neutrino oscillations, and the origin of CP violation, remain to date as open questions in theoretical particle physics.\cite{Giunti:2007ry,Huber:2000ie,PhysRevLett.38.1440,PhysRevLett.44.912,BILENKY1980495,2007NuPhS.167..170M}

In the context of the SM a neutrino flavor created by the weak interaction and 
associated with a charged lepton will maintain its flavor, which implies that 
lepton flavor is conserved and neutrinos are massless. Moreover, recent 
experimental results confirm that neutrinos oscillate, and as 
a consequence, at least two of them have non-zero masses \cite{PhysRevLett.81.1562,PhysRevLett.89.011301,Kajita_2006}.

Current neutrino experiments are measuring the neutrino mixing parameters 
with unprecedented accuracy. The upcoming generation of neutrino 
experiments will be sensitive to subdominant neutrino oscillation effects 
that can in principle give information on the yet-unknown neutrino parameters: 
the Dirac CP-violating phase in the Pontecorvo-Maki-Nakagawa-Sakata (PMNS)
mixing matrix $U_{PMNS}$, the neutrino mass ordering, and the octant of the 
mixing angles. \cite{DUNE:2020ypp,Hyper-Kamiokande:2018ofw,An_2016} 

Up to date, the solar and atmospheric neutrino oscillations have established 
the following values to 3 sigma of the deviation~\cite{10.5281/zenodo.4726908,deSalas:2020pgw,deSalas:2018bym}:
\begin{eqnarray}\label{expn}\nonumber
\Delta m^2_{atm}&=& (2.47 - 2.63) 
\times 10^{-3}\mbox{eV}^2,\\ \nonumber
\Delta m^2_{sol}&=& (6.94 - 8.14) 
\times 10^{-5}\mbox{eV}^2=\Delta m^2_{21}, \\ 
\nonumber
\sin^2\theta_{atm}&=& (4.34 - 6.10) \times 10^{-1} 
=\sin^2\theta_{23},\\ \nonumber
\sin^2\theta_{sol}&=& (2.71-3.69) \times 10^{-1}
=\sin^2\theta_{12},\\ 
\sin^2\theta_{Reac}& = & (2.00-2.41)\times 10^{-2} 
= \sin^2\theta_{13},
\end{eqnarray}
which implies, among other things, that at least two neutrinos have very small 
but non-zero masses.

Masses for neutrinos require physics beyond the SM connected either to the 
existence of right-handed neutrinos and/or to the breaking of the B$-$L 
(baryon minus lepton number) symmetry \cite{PhysRevD.44.3062}. If right-handed neutrinos exist, the 
Yukawa terms lead, after electroweak symmetry breaking, to Dirac neutrino 
masses, requiring Yukawa coupling constants for neutrinos $h_\nu^\phi\leq 10^{-13}$. 
But the right-handed neutrinos, singlets under the SM gauge group, can acquire large Majorana masses and turn the Type I see-saw mechanism \cite{Gell-Mann:1979vob,PhysRevLett.44.912,PhysRevD.22.2227,PhysRevD.23.165} 
an appealing and natural scenery for neutrino mass generation. Another possibility is to generate neutrino masses via quantum loops \cite{Dev:2012sg,Cai:2017jrq}

For Majorana fields, there exists a process called neutrinoless double beta decay $(0\nu\beta\beta)$ which is strongly disfavored by current experimental results
(see~\cite{GERDA2013,CUORE2020,EXO-2002014}), reason for which the alternative is to assume that massive neutrinos must be related to Dirac fields. So, for the model analyzed here, we assume that Majorana masses are forbidden by some kind of physical mechanism.

Besides the fact that no experiment has excluded so far the possibility of Dirac neutrino masses, there are several theoretical motivations to assume them, for example: the generation of baryon asymmetry via leptogenesis~\cite{dick}, alternative approaches to the seesaw mechanism~\cite{wang} and the generation of radiative neutrino masses via quantum loops~\cite{ma1,ma2,ma3,ma4}. Also, in models derived from string theories, the Majorana masses are strongly suppressed by selection rules related to the underlying symmetries~\cite{langa}.

Furthermore, using Dirac particle fields allows us to apply the polar decomposition theorem of algebra \cite{prasolov}, which states that any complex matrix can be decomposed into the product of a Hermitian and a unitary matrix. This decomposition reduces the number of free parameters by half in this sector because the unitary matrix can be absorbed into the singlet representations of $SU(2)_L$; that is, in the right-handed sector (a simplification that is not possible for Majorana particles \cite{Ponce:2011qp}).

Other theoretical motivations to study Dirac neutrinos include the conservation of global lepton number, a common mass generation mechanism for all the Fermion fields, and a clearer distinction between matter and antimatter, which could help to explain CP violation in nature~\cite{GonzalezGarcia2008,Mohapatra2007}. 

To obtain Dirac neutrinos three right-handed neutrinos are added to the  SM of particles and fields (one for each family), allowing in our study for the most general possible Hermitian Dirac mass matrix in the neutral lepton sector. Then, after using a Weak-basis-transformation (WBT) to eliminate nonphysical phases in the hermitian neutral mass matrix, we aim to fit into the parameters the mass-squared differences, and the mixing angles in the $U_{PMNS}$ matrix, values well measured in neutrino physics so far.

In our analysis, we assume a diagonal charged lepton mass matrix in the weak basis, which implies that the mixing angles in $U_{PMNS}$ are pure oscillation parameters with no relation at all with charged lepton mixing.
As a consequence, the unitary matrix that diagonalize the neutral mass matrix is just the same $U_{PMNS}$, and then, by introducing texture zeros in the neutral sector we obtain physical predictions that can be tested numerically.

\section{Zero textures for Dirac Neutrinos}
For the analysis which follows we work with the following three hypotheses:
\begin{enumerate}
 \item We extend the electroweak sector of the SM with three right-handed neutrino fields,  
 $(\nu_{\alpha R};\;\alpha = e,\mu,\tau)$.
 \item The charged lepton mass matrix is diagonal in the weak flavor basis.
 \item Majorana masses are forbidden.
\end{enumerate}
\subsection{Neutrino mass matrix}
According to the previous hypothesis, for the charged lepton sector in the flavor basis, we have
\begin{equation}\label{mslcar}
M_l=\left(\begin{array}{ccc}m_e & 0 & 0 \\ 0 & m_\mu & 0 \\ 0 & 0 & m_\tau 
\end{array}\right),
\end{equation}
which implies that the most general $3\times 3$ mass matrix for the neutrinos, 
which due to the decomposition polar theorem of the matrix algebra \cite{prasolov} we assume hermitian without loss of generality, 
can be written as 
\begin{eqnarray}\nonumber
 M_\nu&=&\left(\begin{array}{ccc} m_{\nu_e\nu_e} & m_{\nu_e\nu_\mu} & m_{\nu_e\nu_\tau} 
\\ m_{\nu_\mu\nu_e} & m_{\nu_\mu\nu_\mu} & m_{\nu_\mu\nu_\tau} 
\\ m_{\nu_\tau\nu_e} & m_{\nu_\tau\nu_\mu} & m_{\nu_\tau\nu_\tau} 
\end{array}\right)= 
 U_{PMNS}\left(\begin{array}{ccc} m_1 & 0 & 0 \\ 0 & m_2 & 0 \\ 0 & 0 & m_3 
\end{array}\right)U_{PMNS}^\dagger \\ \label{mnud}
&=&{\begin{bmatrix}U_{e1}&U_{e2}&U_{e3}\\U_{\mu 1}&U_{\mu 2}&U_{\mu 3}\\
U_{\tau 1}&U_{\tau 2}&U_{\tau 3}\end{bmatrix}}
{\begin{bmatrix}m_1 & 0 & 0 \\ 0 & m_2 & 0 \\ 0 & 0 & m_3 \end{bmatrix}}
{\begin{bmatrix}U_{e1}^*&U_{\mu 1}^*&U_{\tau 1}^*\\U_{e2}^*&U_{\mu 2}^*&U_{\tau 2}^*\\
U_{e3}^*&U_{\mu 3}^*&U_{\tau 3}^*\end{bmatrix}}\\ \nonumber
&=&{\begin{bmatrix}U_{e1}&U_{e2}&U_{e3}\\U_{\mu 1}&U_{\mu 2}&U_{\mu 3}\\
U_{\tau 1}&U_{\tau 2}&U_{\tau 3}\end{bmatrix}}
{\begin{bmatrix}m_1U_{e1}^*&m_1U_{\mu 1}^*&m_1U_{\tau 1}^*\\
m_2U_{e2}^*&m_2U_{\mu 2}^*&m_2U_{\tau 2}^*\\
m_3U_{e3}^*&m_3U_{\mu 3}^*&m_3U_{\tau 3}^*\end{bmatrix}}
\end{eqnarray}
where the mixing matrix $U_{PMNS}$ for Dirac neutrinos is parametrized in 
the usual way as \cite{ParticleDataGroup:2020ssz}:
\begin{equation}\label{mnudp}
{\displaystyle{\begin{aligned}&{U_{PMNS}=\begin{bmatrix}1&0&0\\
0&c_{23}&s_{23}\\0&-s_{23}&c_{23}\end{bmatrix}}
{\begin{bmatrix}c_{13}&0&s_{13}e^{-i\delta _{\text{CP}}}\\
0&1&0\\-s_{13}e^{i\delta _{\text{CP}}}&0&c_{13}\end{bmatrix}}
{\begin{bmatrix}c_{12}&s_{12}&0\\-s_{12}&c_{12}&0\\0&0&1\end{bmatrix}}\\
&={\begin{bmatrix}c_{12}c_{13}&s_{12}c_{13}&s_{13}e^{-i\delta _{\text{CP}}}\\
-s_{12}c_{23}-c_{12}s_{23}s_{13}e^{i\delta _{\text{CP}}}&
c_{12}c_{23}-s_{12}s_{23}s_{13}e^{i\delta _{\text{CP}}}&
s_{23}c_{13}\\s_{12}s_{23}-c_{12}c_{23}s_{13}e^{i\delta _{\text{CP}}}&
-c_{12}s_{23}-s_{12}c_{23}s_{13}e^{i\delta _{\text{CP}}}&c_{23}c_{13}\end{bmatrix}};
\end{aligned}}} 
\end{equation}
where $Dg.(m_1,m_2,m_3)$ refers to the neutrino mass eigenvalues, and 
$c_{ij}=\cos\theta_{ij}$ and $s_{ij}=\sin\theta_{ij}$ are the cosine and sine 
of the mixing angle $\theta_{ij},\;\;i < j=1,2,3$.

Now, due to the hermiticity constraint, the elements of $M_\nu$ satisfy:
$m_{\nu_e\nu_e}=m_{\nu_e\nu_e}^*,\; 
m_{\nu_\mu\nu_\mu}=m_{\nu_\mu\nu_\mu}^*,\; 
m_{\nu_\tau\nu_\tau}=m_{\nu_\tau\nu_\tau}^*$, and 
$m_{\nu_\mu\nu_e}=m_{\nu_e\nu_\mu}^*$; 
$m_{\nu_\tau\nu_e}=m_{\nu_e\nu_\tau}^*$ and 
$m_{\nu_\mu\nu_\tau}=m_{\nu_\tau\nu_\mu}^*$.

For our analysis, it is convenient to use the following numerical values 
for the entries of $U_{PMNS}$ evaluated at $3\sigma$ ranges, 
presented in the literature~\cite{10.5281/zenodo.4726908}:

\begin{equation}\label{pmnN}
\left(\begin{array}{ccc}
         0.7838...0.8442 & 0.5135...0.6004 & 0.1901...0.2183 \\
         0.2508...0.4902 & 0.4665...0.6782 & 0.6499...0.7719 \\
         0.3135...0.5471 & 0.4841...0.6927 & 0.6161...0.7434 \\
        \end{array} \right),
\end{equation}
numbers which include strong correlations between the allowed ranges do to 
unitary constraints.

\subsubsection{Counting parameters}
When the mass matrices for the lepton sector are given by 
(\ref{mslcar}) and (\ref{mnud}), we have that the hermitian mass matrix 
$M_\nu$ has six real parameters and three phases that we can use to 
explain seven physical parameters: the three mixing angles 
$\theta_{12},\;\;\theta_{13}$ and $\theta_{23}$, one CP violating phase $\delta$, 
and three neutrino masses  $m_1,\; m_2$ and $m_3$. So, in principle, we have a redundant number of parameters (two more phases).

Now, at this point, and contrary to the quark sector \cite{Ponce:2013nsa,Ponce:2011qp}, we can not introduce texture zeros via WBT~\cite{FRITZSCH20001,branco2009,PhysRevD.87.053016,Benavides:2022hca} in the mass matrix $M_\nu$ due to the fact that it will change the charged lepton diagonal mass matrix. But as it is shown in the appendix, the ``Weak basis transformations'' can be used to eliminate the two redundant phases. 

Once the redundant phases are removed via WBT, the hermitian matrix $M_\nu^\prime$ ends up with six real parameters and one phase able to accommodate, in principle, the three mixing angles, the three neutrino masses, and the CP violation phase. So, one texture zero should imply a relationship between the mixing angles and the physical masses.

Unfortunately, we do not have six experimental entries to input in the analysis, because the neutrinos masses are not known. What we know instead are the mass square differences 
$\Delta m_{32}^2=m_3^2-m_2^2$; $\Delta^2_{31}=m_3^2-m_1^2$, and $\Delta m_{21}^2=m_2^2-m_1^2$ in normal hierarchy, with the mathematical constraint $\Delta m_{21}^2+\Delta m^2_{32}-\Delta m^2_{31}=0$ which leave us with only five experimental real constraints to be accommodated. So, patterns with one texture zero should, in principle, be compatible with the experimental data at the $3\sigma$ level, although the parameter space, for each of these zero textures, should be strictly constrained (an analysis presented somewhere else). So, real physical predictions should start only when two texture zeros are considered. 

\subsection{Texture Zeros}
Introducing texture zeros in a general mass matrix has been an outstanding hypothesis that provides relationships between the mixing angles and the mass values.

As discussed above, the six real mathematical parameters of the most general hermitian mass matrix for the case of Dirac neutrinos provides enough room to accommodate the five real experimental values with no prediction at all. Even further, one texture zero should not conduce to any prediction either. So, texture zeros start to be worth when two of them are introduced, with two texture zeros providing one physical prediction.

In what follows we will study, for the case of normal ordering, all the possible cases of two texture zeros in the hermitian mass matrix for Dirac neutrinos, and see what the prediction for the lightest Dirac neutrino mass is, which has, as a consequence, the knowledge of the complete neutrino mass spectrum.

Three different cases must be analyzed: two texture zeros in the diagonal, one texture zero in the diagonal and the other outside the diagonal, and finally two off-diagonal texture zeros.

To set our mathematical notation, let us start studying the implications of one texture zero.

\subsubsection{Diagonal texture zeros}
Let us start assuming that $m_{\nu_e\nu_e}=0$ and see its implications: 

From equation (\ref{mnud}) we have: 
\begin{equation}\label{tz12d}
 m_{\nu_e\nu_e}=m_1|U_{e1}|^2 +m_2|U_{e2}|^2 +m_3|U_{e3}|^2 =0,
\end{equation}
dividing by $m_3$ and using the unitary constraint of matrix  $U$; that is 
$|U_{e1}|^2 +|U_{e2}|^2 +|U_{e3}|^2=1$ we can writte (\ref{tz12d}) as: 
\[\frac{m_1}{m_3}|U_{e1}|^2+\frac{m_2}{m_3}|U_{e2}|^2+1-|U_{e1}|^2-|U_{e2}|^2=0;\]
which we can rearrenge as:
\begin{equation}\label{mnudp1}
 |U_{e2}|^2=\frac{m_3}{m_3-m_2}-\frac{m_3-m_1}{m_3-m_2}|U_{e1}|^2.
\end{equation}

In a similar way for $m_{\nu_\mu\nu_\mu}=0$ we have:
\begin{equation}\label{mnudp2}
 |U_{\mu 2}|^2=\frac{m_3}{m_3-m_2}-\frac{m_3-m_1}{m_3-m_2}|U_{\mu 1}|^2,
\end{equation}
and for $m_{\nu_\tau\nu_\tau}=0$ we have:
\begin{equation}\label{mnudp2}
 |U_{\tau 2}|^2=\frac{m_3}{m_3-m_2}-\frac{m_3-m_1}{m_3-m_2}|U_{\tau 1}|^2.
\end{equation}

The former three cases can be summarized as:
\begin{equation}\label{mnudpg}
 |U_{\alpha 2}|^2=\frac{m_3}{m_3-m_2}-\frac{m_3-m_1}{m_3-m_2}|U_{\alpha 1}|^2, 
\end{equation}
for $\alpha=e$ if $m_{\nu_e\nu_e}=0$; $\alpha=\mu$ if $m_{\nu_\mu\nu_\mu}=0$, 
and $\alpha=\tau$ if $m_{\nu_\tau\nu_\tau}=0$. This shows the dependence between two of the $U_{PMNS}$ matrix entries and the neutrino mass values.

\subsubsection{Texture zeros outside the diagonal}
Let us consider now a texture zero outside the diagonal. Let us start with 
$m_{\nu_e\nu_\mu}=0$ (notice that $m_{\nu_\mu\nu_e}=m_{\nu_e\nu_\mu}^*=0$).

For this situation equation (\ref{mnud}) implies that: 
\begin{equation}\label{tz12nd}
 m_{\nu_e\nu_\mu}=m_1U_{e1}U_{\mu 1}^* +m_2U_{e2}U_{\mu 2}^* +m_3U_{e3}U_{\mu 3}^* =0,
\end{equation}
which dividing by $m_3$ and using the orthogonality condition 
$U_{e1}U_{\mu 1}^* +U_{e2}U_{\mu 2}^* +U_{e3}U_{\mu 3}^*=0$ can be written as: 

\begin{equation}\label{tz12p}
 \left(\frac{m_1}{m_3}-1\right)U_{e1}U_{\mu 1}^* +
\left(\frac{m_2}{m_3}-1\right)U_{e2}U_{\mu 2}^*=0,
\end{equation}
multiplying by $U_{e2}^*U_{\mu 2}$ and rearranging we have 
\begin{equation}\label{tz12pp}
 \left(\frac{m_1}{m_3}-1\right)U_{e1}U_{\mu 1}^*U_{e2}^*U_{\mu 2} +
\left(\frac{m_2}{m_3}-1\right)|U_{e2}|^2|U_{\mu 2}|^2=0, 
\end{equation}
which we can finally write as
\begin{equation}\label{tz12ppp}
U_{e1}U_{\mu 1}^*U_{e2}^*U_{\mu 2} +
\left(\frac{m_3-m_2}{m_3-m_1}\right)|U_{e2}|^2|U_{\mu 2}|^2=0, 
\end{equation}
equation, that together with its complex conjugate, can be separated in two parts: 
a real part equal to zero and an imaginary part also equal to zero (notice that 
for a hermitian matrix its eigenvalues must be real but not necessarily positive).

As $m_{\nu_\mu\nu_e}$ must also be equal to zero, the two relations must also be 
equivalent to taking the real part and the imaginary part in (\ref{tz12ppp})
equal to zero. As the second term in (\ref{tz12ppp}) is real, to take the 
imaginary part equal to zero produces 
\begin{equation}
 Im.(U_{e1}U_{\mu 1}^*U_{e2}^*U_{\mu 2})=J=0; 
\end{equation}
which means that this texture zero is associated with a Jarlskog invariant \cite{PhysRevLett.55.1039} 
equal to zero and no CP violation is present for this texture zero.

In a similar way for $m_{\nu_e\nu_\tau}=0$, we have: 
\begin{equation}\label{tz13nd}
 m_{\nu_e\nu_\tau}=m_1U_{e1}U_{\tau 1}^* +m_2U_{e2}U_{\tau 2}^* + 
 m_3U_{e3}U_{\tau 3}^* =0.
\end{equation}
Dividing by $m_3$ and using the orthogonality relationship 
$U_{e1}U_{\tau 1}^* +U_{e2}U_{\tau 2}^* +U_{e3}U_{\tau 3}^*=0$ we can write  
(\ref{tz13nd}) in the form: 
\begin{equation}\label{tz13p}
 \left(\frac{m_1}{m_3}-1\right)U_{e1}U_{\tau 1}^* +
\left(\frac{m_2}{m_3}-1\right)U_{e2}U_{\tau 2}^*=0,
\end{equation}
multiplying by $U_{e2}^*U_{\tau 2}$ and rearranging, we have 
\begin{equation}\label{tz13pp}
 \left(\frac{m_1}{m_3}-1\right)U_{e1}U_{\tau 1}^*U_{e2}^*U_{\tau 2} +
\left(\frac{m_2}{m_3}-1\right)|U_{e2}|^2|U_{\tau 2}|^2=0, 
\end{equation}
which in turn implies 
\begin{equation}\label{tz13ppp}
U_{e1}U_{\tau 1}^*U_{e2}^*U_{\tau 2} +
\left(\frac{m_3-m_2}{m_3-m_1}\right)|U_{e2}|^2|U_{\tau 2}|^2=0, 
\end{equation}
which again produces  
\begin{equation}
 Im.(U_{e1}U_{\tau 1}^*U_{e2}^*U_{\tau 2})=J=0. 
\end{equation}

The former means that this texture zero outside the diagonal is also associated with a Jarlskog invariant equal to zero and 
again, there is no CP violation for this case.

In a similar way we have for $m_{\nu_\mu\nu_\tau}=0$ that 
\begin{equation}\label{tz23nd}
 m_{\nu_\mu\nu_\tau}=m_1U_{\mu 1}U_{\tau 1}^* +m_2U_{\mu 2}U_{\tau 2}^* + 
 m_3U_{\mu 3}U_{\tau 3}^* =0,
\end{equation}
which divided by $m_3$ and making use of the appropriate orthogonality relationship, 
we have
\begin{equation}\label{tz23p}
 \left(\frac{m_1}{m_3}-1\right)U_{\mu 1}U_{\tau 1}^* +
\left(\frac{m_2}{m_3}-1\right)U_{\mu 2}U_{\tau 2}^*=0,
\end{equation}
which multiplied by $U_{\mu 2}^*U_{\tau 2}$ we get  
\begin{equation}\label{tz23ppp}
U_{\mu 1}U_{\tau 1}^*U_{\mu 2}^*U_{\tau 2} +
\left(\frac{m_3-m_2}{m_3-m_1}\right)|U_{\mu 2}|^2|U_{\tau 2}|^2=0, 
\end{equation}
which again takes to 
\begin{equation}
 Im.(U_{\mu 1}U_{\tau 1}^*U_{\mu 2}^*U_{\tau 2})=J=0; 
\end{equation}

So, a texture zero in $M_\nu$ outside the main diagonal implies CP conservation, a result also obtained in a different way in appendix A.

\section{Numerical analysis}
According to equation~(\ref{mnudp}),  the right-hand side of equation~(\ref{mnud}) depends only on the neutrino mixing angles, the CP-violating phase, and the neutrino masses. So, each one of the possible six texture zeros in the matrix $M_\nu$ in equation~(\ref{mnud}) will imply an equation relating neutrino masses with the CP-phase and the neutrino mixing angles. Equation that must be confronted with the measured experimental values. To do it we must put each equation in terms of physical parameters. Let us see:

\subsection{$m_{\nu_e\nu_e}=0$}
The texture $m_{\nu_e\nu_e}=0$ produce the constraint in equation (\ref{mnudp1}) 
which when put in terms of physical parameters becomes:
\begin{equation}\label{one}
s_{12}^2c_{13}^2=\frac{m_3}{(m_3-m_2)}-\frac{(m_3-m_1)}{(m_3-m_2)}c_{12}^2c_{13}^2.
\end{equation}

Relationship that must be satisfied by the experimental measured values in order to have a realistic texture zero in the neutrino mass matrix.

The relationship (\ref{one}) can be rearranged a little by using the definitions 
$m=m_1+m_2+m_3$ and $\Delta m_{32}^2=m_3^2-m_2^2$

\begin{equation}\label{oner}
\frac{s_{12}^2c_{13}^2\Delta m_{32}^2}{(m-m_1)}=
m_3-(m_3-m_1)c_{12}^2c_{13}^2.
\end{equation}

The parameter space can be studied through a $\chi^2$ analysis, which is defined as

\begin{equation}\label{chi2}
\chi^2(m_1)= \left( \frac{\sin^2\theta_{12}-\sin^2\tilde{\theta}_{12}}{\sigma(\sin^2\theta_{12})} \right)^2  
\end{equation}

\noindent
where $\sin^2\tilde\theta_{12}$ is the value for this mixing angle  obtained from Eq. (\ref{oner}),
while $\sin^2\theta_{12}$  and $\sigma(\sin^{2}\theta_{12})$ are the current best fit value and its one sigma deviation, respectively.
Experimental data in (\ref{expn}) and (\ref{pmnN}) are used in order to carry the analysis.

After defining the parameter space in terms of $\Delta m_{12}^{2}$, $\Delta m_{13}^{2}$ and the mixing angles $\theta_{12}$, $\theta_{13}$,   we perform a minimization process for the $\chi^2$ function.  The best fit points obtained around the minimum (about zero) for our analysis were:
$$m_1=0.00208\,\, \text{eV}, \quad m_2=-0.00886 \,\,\text{eV},\quad m_3=0.0501\,\,\text{eV}$$  

\begin{figure}[h!]\label{fig1}
     \centering
     \begin{subfigure}[b]{0.3\textwidth}
         \centering
         \includegraphics[width=\textwidth]{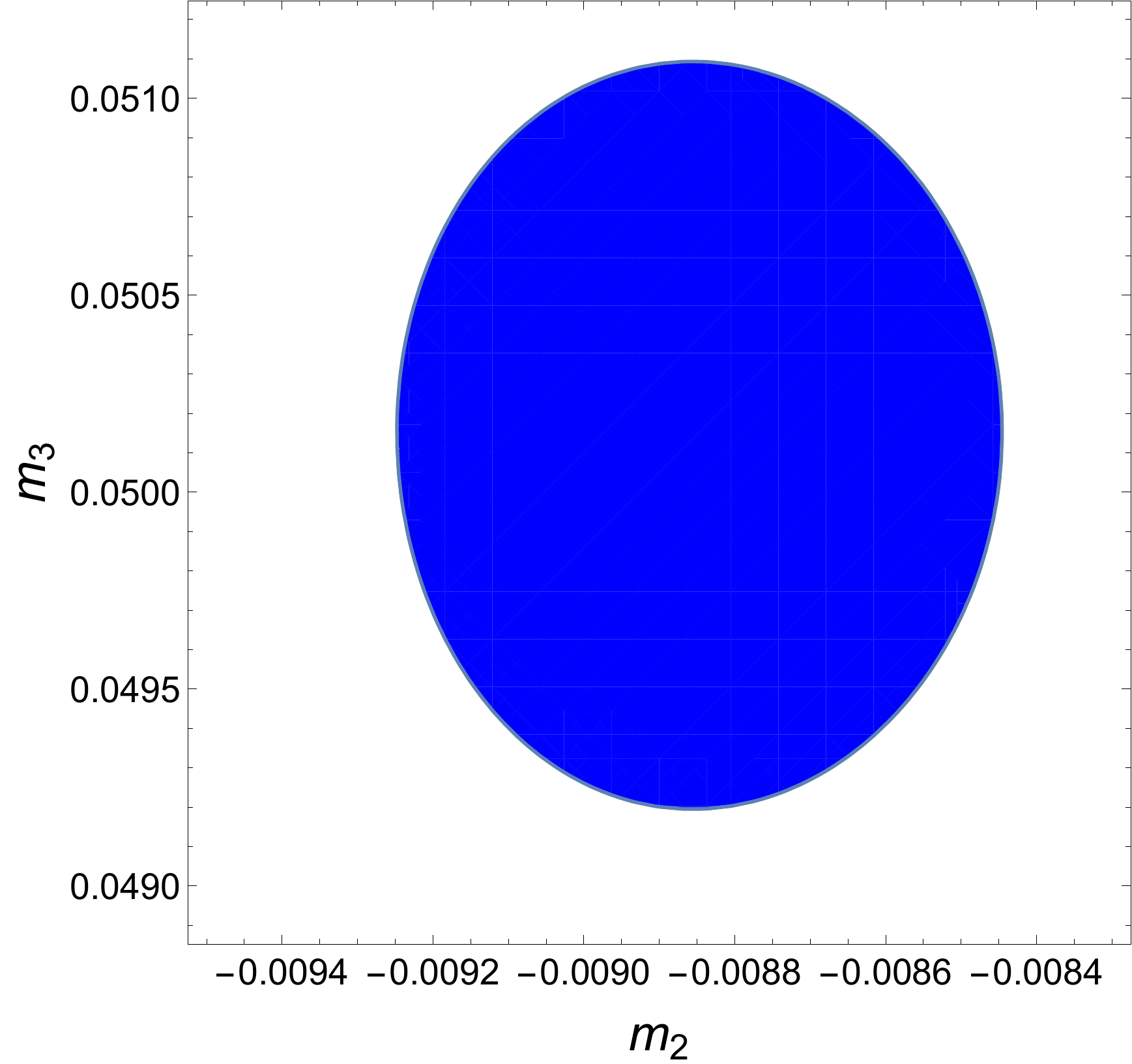}
         \caption{Plot evaluated at $m_1=0.00208$ eV.}
         \label{fig:fijom1}
     \end{subfigure}
     \hfill
     \begin{subfigure}[b]{0.3\textwidth}
         \centering
         \includegraphics[width=\textwidth]{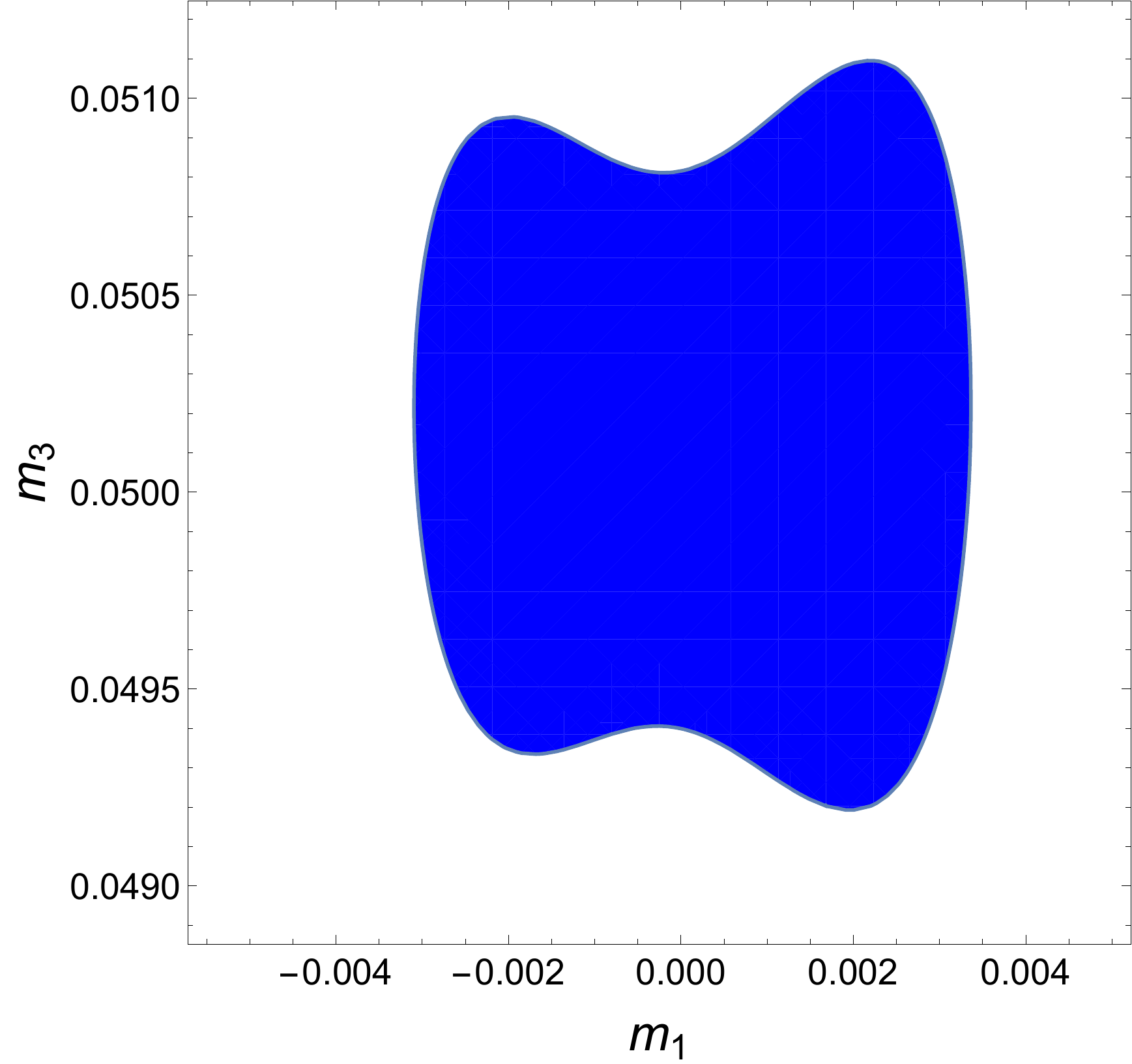}
         \caption{Plot evaluated at $m_2=-0.00886$ eV.}
         \label{fig:fijom2}
     \end{subfigure}
     \hfill
     \begin{subfigure}[b]{0.3\textwidth}
         \centering
         \includegraphics[width=\textwidth]{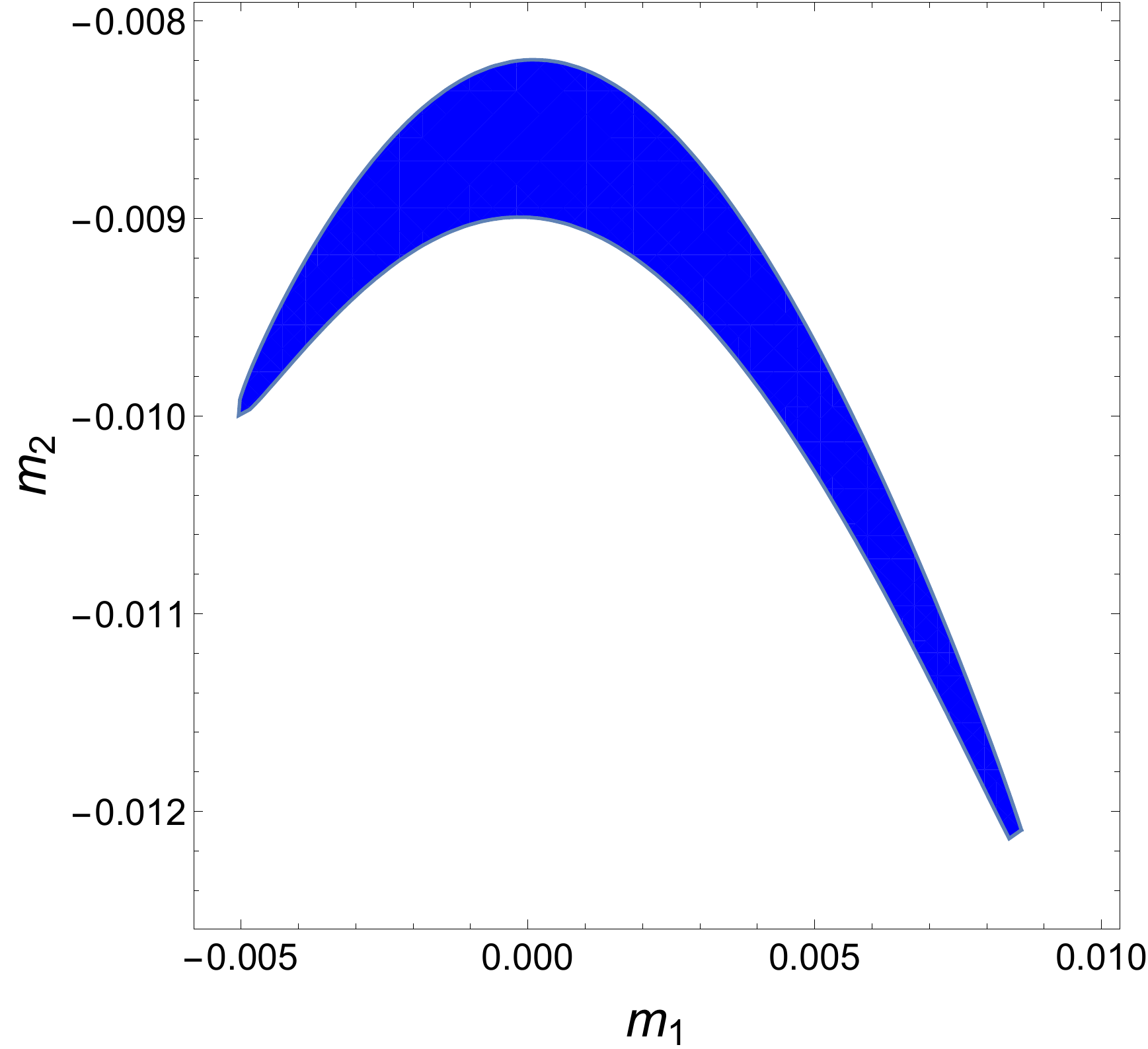}
         \caption{Plot evaluated at $m_3=0.0501$ eV.}
         \label{fig:fijom3}
     \end{subfigure}
        \caption{These figures presents the parameter space  for the texture zero $M_{\nu_e \nu_e}=0$ at 95\% C.L.}
        \label{fig:1ZT}
\end{figure}

In order to study the $(m_1,\,m_2,\,m_3)$ parameter space, we fix first the $m_1$ value in its minimum and look for the allowed values for $m_2$ and $m_3$ at $95\%$ confidence level (CL), results presented in the figure (\ref{fig:fijom1}). Next we fix $m_2$ in its minimum and look for the allowed values for $m_1$ and $m_3$ at 95 CL, 
results presented in (\ref{fig:fijom2}). Finally, fixing $m_3$ we find the parameter space allowed for $m_1$ and $m_2$ as presented in 
fig. (\ref{fig:fijom3}).  The parameter spaces shown in the former three figures satisfies the experimental limits $|m_1|+|m_2|+|m_3|<0.12$ $eV$ \cite{refId0}, and also the square mass differences and the three mixing angles.

From Eq.~(\ref{one}) we can see that the CP violation phase is entirely unconstrained, but the plots in the pictures show that the experimental measured values can be well accommodated in the allowed parameter space, as anticipated above. The mixing angle $\theta _{23}$ is easily obtained from the $ (2,3)$  or $ (3,3)$  mixing matrix numerical entries.

A similar analysis can be carried through for the other five one texture zero in the matrix $M_\nu$. That is, for $m_{\nu_\mu\nu_\mu}=0$, $m_{\nu_\tau\nu_\tau}=0$, $m_{\nu_e\nu_\mu}$
$m_{\nu_\mu\nu_e}=0$, etc. The results for this analysis will be presented elsewhere. 

\subsection{Two texture zeros} \label{FiveTZ} 
The next step is to study the different structures with two texture zeros in the neutrino mass matrix (with a diagonal charged lepton sector in the weak basis). There are three different cases: first, the two zeros are in the main diagonal (there are three CP violating patterns); next, one texture zero is in the main diagonal and the other one outside this diagonal (with nine CP conserving different patterns); and, finally, the two zeros are off the main diagonal (with three CP conserving different patterns).\footnote{Other possibilities with two texture zeros in the neutral sector and three texture zeros in the charged sector with at least one of them in the diagonal are analyzed in refs. \cite{Benavides:2022hca,Rico:2023drr,Benavides:2020pjx} }  

In the following section we will present, for one particular pattern, the detailed analytic and numerical analysis we have carried through for all the different fifteen two texture zeros patterns. Our numerical results are presented in one appendix at the end of the paper.


\subsubsection{Two zeros, one of them in the main diagonal}\label{sec:TwoZeros}
The number of different patterns in this category, all of them related to CP conservation are nine. They are shown in  Appendix 6.2.1.

In our analysis, carried through in two steps, we reconstruct first the neutrino mass matrix in terms of the three neutrino masses $m_1,\;\; m_2$ and $m_3$. This is achieved by making use of the invariant forms: tr$[M]$, tr$[M^2]$, and det$[M]$. After that we derive the analytic orthogonal matrices that diagonalize the several 3×3  real neutrino mass matrices. The result is the $U_{PMNS}$ in analytic form as a function of the real mass eigenvalues and any other parameter needed, this last one conveniently chosen. 

To see this, let us take as an example here the texture in $A_7$: 

\begin{equation}
   A_7= \begin{pmatrix*}[c]
        x_1  &  0   &  b\\
        0    & x_2  &  c\\
        b^{*} & c^{*} &  0
    \end{pmatrix*}.
\end{equation}

Notice that the three eigenvalues of a general $3\times 3$ Hermitian matrix are real, but not necessarily positive. As a mater of fact, taking the determinant of $A_7$ we have $|A_7|=-x_2|b|^2-x_1|c|^2=m_1m_2m_3$ which by fixing $m_3>0$ by a global phase convention, we must have two classes of solutions: $m_1>0,\;\; m_2<0$ and $m_1<0,\;\;\; m_2>0$. 
Let us carry through our example for the particular case $m_1 > 0$ and  $m_2<0$

After using the invariant forms for this $3\times 3$ mass matrix, and solving the equations, we have: 

\begin{equation}
\begin{pmatrix*}[c]
 x_1 & 0 & \sqrt{\frac{\left(x_1-m_1\right) \left(x_1+m_2\right) \left(-x_1+m_3\right)}{2 x_1-m_1+m_2-m_3}} \\
 0 & -x_1+m_1-m_2+m_3 & \sqrt{\frac{\left(x_1-m_1+m_2\right) \left(-x_1+m_1+m_3\right) \left(x_1+m_2-m_3\right)}{2 x_1-m_1+m_2-m_3}} \\
 \sqrt{\frac{\left(x_1-m_1\right) \left(x_1+m_2\right) \left(-x_1+m_3\right)}{2 x_1-m_1+m_2-m_3}} & \sqrt{\frac{\left(x_1-m_1+m_2\right) \left(-x_1+m_1+m_3\right) \left(x_1+m_2-m_3\right)}{2 x_1-m_1+m_2-m_3}} & 0 \\
\end{pmatrix*}; 
\end{equation}
where $m_1,\,m_2,\,m_3$ and $x_1$ are free parameters used to calculate the neutrino masses and the mixing angles. After diagonalizing this texture, we obtained the $U_{\text{PMNS}}$ in terms of the free parameter.

\resizebox{\linewidth}{!}{%
$\displaystyle
U_{\text{PMNS}}= 
\left( \begin{array}{rrr} 
-\sqrt{\frac{\left(x_1+m_2\right) \left(x_1+m_2-m_3\right) \left(m_3-x_1\right)}{\left(m_1+m_2\right) \left(m_3-m_1\right) \left(2 x_1-m_1+m_2-m_3\right)}} & \sqrt{\frac{\left(x_1-m_1\right) \left(x_1-m_1+m_2\right) \left(-x_1+m_1+m_3\right)}{\left(m_1+m_2\right) \left(m_1-m_3\right) \left(-2 x_1+m_1-m_2+m_3\right)}} & \sqrt{\frac{\left(x_1-m_1\right) \left(x_1+m_2-m_3\right)}{\left(m_1+m_2\right) \left(-m_1+m_3\right)}} \\
 -\sqrt{\frac{\left(m_1-x_1\right) \left(m_3-x_1\right) \left(-x_1+m_1+m_3\right)}{\left(m_1+m_2\right) \left(m_2+m_3\right) \left(-2 x_1+m_1-m_2+m_3\right)}} & -\sqrt{\frac{\left(x_1+m_2\right) \left(x_1-m_1+m_2\right) \left(x_1+m_2-m_3\right)}{\left(m_1+m_2\right) \left(m_2+m_3\right) \left(2 x_1-m_1+m_2-m_3\right)}} & \sqrt{\frac{\left(x_1+m_2\right) \left(-x_1+m_1+m_3\right)}{\left(m_1+m_2\right) \left(m_2+m_3\right)}} \\
 \sqrt{\frac{\left(x_1-m_1\right) \left(x_1+m_2\right) \left(x_1-m_1+m_2\right)}{\left(m_1-m_3\right) \left(m_2+m_3\right) \left(-2 x_1+m_1-m_2+m_3\right)}} & \sqrt{\frac{\left(x_1-m_3\right) \left(x_1-m_1-m_3\right) \left(x_1+m_2-m_3\right)}{\left(m_1-m_3\right) \left(m_2+m_3\right) \left(-2 x_1+m_1-m_2+m_3\right)}} & \sqrt{\frac{\left(x_1-m_1+m_2\right) \left(x_1-m_3\right)}{\left(m_1-m_3\right) \left(m_2+m_3\right)}} \\
\end{array} \right)
$}
\vspace{1cm}

Once we obtained this expression we made a $\chi^2$ minimization procedure 

\begin{equation}
\chi^2(m_1,x_1)= \sum_{i<j}\left( \frac{\sin^2\theta_{ij}-\sin^2\tilde{\theta}_{ij}}{\sigma(\sin^2\theta_{ij})} \right)^2 ,\,\,\,\text{with} \,\,\, i,j=1,2,3. 
\label{eq:xi2}
\end{equation}

\noindent
where $\sin^2\tilde\theta_{ij}$ are the angles mixing predicted by our forms, 
while $\sin^2\theta_{ij}$  and $\sigma(\sin^{2}\theta_{ij})$ are the current best-fit values and its one sigma deviation, respectively.  To obtain the best values that fit the experimental data. Notice that the analytical results contain some square root terms which imply several limits on the parameters, such that the results are real; that is, we must have
$$m_3>x_1>m_1,\;2x_1>m_3,\;x_1+m_2>m_3$$

The minimization procedure left the following phenomenological results: The neutrino masses $|m_1|=0.0333\,\,\,$ eV, $\quad |m_2|=0.0344\,\,\,$ eV, $\quad |m_3|=0.0608 \,\,$ eV, and the mixing angles $\sin^2{\theta_{12}}=0.315,\quad\sin^2{\theta_{23}}=0.646,\quad \sin^2{\theta_{13}}=0.022.$



Numerical result in agreement with the data reported experimentally by the Neutrino Global Fit\cite{10.5281/zenodo.4726908}.

The results of the other eight possibilities are shown in Appendix D.

\vspace{0.5cm}

\subsubsection{Two zeros off the main diagonal}
None of the three cases is viable because each one of them is associated with a vanishing oscillation parameter (for $A_{10}$ we have $\theta_{13}=0$, for $A_{11}$ we have $\theta_{23}=0$ and for $A_{12}$ we have $\theta_{12}=0$), The forms are showed in Appendix: \ref{dosceros}.

{\subsubsection{Two zeros in the main diagonal}}
There are three different patterns given by the matrices $A_{13},\; A_{14}$ and $A_{15}$. Using the invariant forms as before we end up with analytic expressions for the $U_{PMNS}$ matrix quite complicated, whose tracking is not much illuminating. So we proceed immediately with the numerical analysis.

Our result shows that none of the three different textures with two zeroes in the main diagonal is able to reproduce the three measured mixing angles in the $U_{PMNS}$ oscillation matrix.

\section{Summary}
In the context of a model with right-handed neutrinos (one for each family) and global lepton number conservation, we have performed an  analytic and numerical systematic study of the Dirac neutrino Hermitian mass matrix $M_\nu$ with two independent texture zeros, under the assumption that the charged lepton sector is diagonal in the weak basis.

Analytic expressions for the entries of $M_\nu$ as functions of the the three neutrino masses are obtained by using the mathematical invariant of a $3\times 3$ matrix. Algebraic expressions are derived to obtain numerical values for the physical parameters via minimization with a $\chi^2$ statistical analysis. 

According to our study, the cases compatible with the current experimental data at the $3\sigma$ level are  $A_3$ and $A_7$ (in appendix B), both of them associated to normal ordering an CP conservation. This is contrary to the results presented in  Refs.~\cite{lenis2023twozero,liu} where the analysis was carried through but looking for correlations between two of the three mixing angles. Now, performing our analysis but relaxing the correlation between the angle  $\sin{\theta_{13}}$ with the other two mixing parameters at the $3\sigma$ level, the results in Table [3] are obtained, now in agreement with the results reported in the literature ~\cite{lenis2023twozero,liu}.

A new feature in our analysis is the iterative use of weak basis transformations~\cite{FRITZSCH20001,branco2009,PhysRevD.87.053016,Benavides:2022hca} which provided us with: first, the elimination of the two redundant phases in the most general hermitian neutrino mass matrix ending up with only one physical phase connected with possible CP violation in he lepton sector; and second, the demonstration in a novel way the CP conservation in the context of our analysis when a texture zero outside the main diagonal is placed.

We want to stress that in our analysis, based on the assumption of a diagonal charged lepton mass matrix in the weak basis, our $U_{PMNS}$ matrix is a pure oscillation matrix, and not a mixing one as it occurs in the quark sector. This is relevant because oscillations is what has been measured in the neutrino experiments.

From our study the mass for one of the three neutrinos can be predicted (we choose the lightest one). Using this value, and the experimental mass squared differences, the entire neutrino mass spectrum can be inferred, as presented in the three tables at the end of the paper. Those values are exact predictions in our analysis. For example, for $A_3$ in Table I we have that $m_1=0.021$, $m_2=0.087$, and $m_3=0.501$, values predicted in eV.

Although we realized that all the six one zero textures are compatible with current neutrino oscillations, we have not carried through in detail the constraints in the parameter space coming from the experimental measured values (results presented elsewhere).

Last but not least, it is worthwhile to stress that whether neutrinos are Dirac or Majorana particles, remains an open question. 

\section*{Acknowledges}
 R. H. B. acknowledges additional financial support from Minciencias CD82315 CT ICETEX 2021-1080.

\section{Appendix A}\label{appendixA}
In this appendix we address two issues: first, we show how to use the weak basis transformation in order to reduce the number of phases from three to one in a general $3\times 3$ hermitian neutrino mass matrix (for the case of a diagonal mass matrix in the charged lepton sector). Second, we discuss the mathematical reason for CP conservation when there is an off-diagonal vanishing element in the neutrino sector.

In the context of the SM extended with right-handed neutrinos and lepton number conservation, the most general 
weak basis transformation that leaves the two $3\times 3$ lepton mass 
matrices Hermitian, and does not alter the physics implicit in the weak currents 
(does not alter the physical content in the $U_{PMNS}$  mixing matrix), is an 
arbitrary unitary transformation $U$ acting simultaneously in the charged lepton 
and in the neutrino mass matrices \cite{branco2009}. That is
\begin{equation}\label{2aa}
\begin{split}
M_\nu&\longrightarrow M_\nu^R=U M_\nu U^\dag,\\
M_l&\longrightarrow M_l^R=U M_l U^\dag.
\end{split}
\end{equation}

Now, when the mass matrices for the charged lepton sector are diagonal,  
we have that the most general hermitian mass matrix $M_\nu$  for the neutral sector has six real parameters and three phases that we can use to explain seven physical parameters: three neutrino masses  $m_1, m_2$ and $m_3$, the three mixing angles $\theta_{12}, \theta_{13}$ and $\theta_{23}$,
and one CP violating phase $\delta$ in the $U_{PMNS}$ mixing matrix. So, in principle, 
we have a redundant number of parameters (two more phases).

Contrary to the quark sector~\cite{Ponce:2013nsa,Ponce:2011qp}, we can not introduce texture zeros via weak 
basis transformations in the mass matrix $M_\nu$ because 
it would change the charged lepton diagonal mass matrix. However, the ``Weak basis transformations'' can eliminate 
the redundant phases. To do this, let us write the neutrino mass matrix as:

\begin{equation}
M_{\nu}=\begin{pmatrix*}[l]\label{mngenp}
\vert m_{\nu_e\nu_e} \vert && \vert m_{\nu_e\nu_\mu} \vert e^{i\phi_{e\mu}} && \vert m_{\nu_e\nu_\tau} \vert e^{i\phi_{e\tau}} \\
\vert m_{\nu_e\nu_\mu} \vert e^{-i\phi_{e\mu}} && \vert m_{\nu_\mu\nu_\mu} \vert && \vert m_{\nu_\mu\nu_\tau} \vert e^{i\phi_{\mu\tau}} \\
\vert m_{\nu_e\nu_\tau} \vert e^{-i\phi_{e\tau}} && \vert m_{\nu_\mu\nu_\tau} \vert e^{-i\phi_{\mu\tau}} && \vert m_{\nu_\tau\nu_\tau} \vert
\end{pmatrix*};
\end{equation}

and let us do a weak basis transformation using the following diagonal unitary 
matrix:
\[M_\phi=\text{Diag}(e^{i\phi_1},1,e^{i\phi_2}),\;\;\; 
M_\phi^\dagger=\text{Diag}(e^{-i\phi_1},1,e^{-i\phi_2})=M_\phi^{-1},\]
which does not change the diagonal charged lepton mass matrix. 
Afther this, the matrix (\ref{mngenp}) gets the following form:

\begin{equation}\nonumber
M_\nu^\prime= \begin{pmatrix*}[l]
|m_{\nu_e\nu_e}| && |m_{\nu_e\nu_\mu}|e^{i(\phi_{e\mu}-\phi_1)} && |m_{\nu_e\nu_\tau}|e^{i(\phi_{e\tau}+\phi_2-\phi_1)} \\ 
|m_{\nu_e\nu_\mu}|e^{-i(\phi_{e\mu}-\phi_1)} && |m_{\nu_\mu\nu_\mu}| && |m_{\nu_\mu\nu_\tau}|e^{i(\phi_{\mu\tau}+\phi_2)} \\ 
|m_{\nu_e\nu_\tau}|e^{-i(\phi_{e\tau}+\phi_2-\phi_1)} && |m_{\nu_\mu\nu_\tau}|e^{-i(\phi_{\mu\tau}+\phi_2)} && |m_{\nu_\tau\nu_\tau}| 
\end{pmatrix*}; 
\end{equation}

where $M_\nu^\prime=M_\phi^\dagger M_\nu M_\phi$. Three cases are present in this 
expression:\\

{\bf Case A}: $\phi_1=\phi_{e\mu}$ and 
$\phi_2=\phi_1-\phi_{e\tau}=\phi_{e\mu}-\phi_{e\tau}$. Producing
\begin{equation}\label{mngepA}
M_\nu^\prime=\begin{pmatrix*}[l]
|m_{\nu_e\nu_e}| && |m_{\nu_e\nu_\mu}| && |m_{\nu_e\nu_\tau}| \\ 
|m_{\nu_e\nu_\mu}| && |m_{\nu_\mu\nu_\mu}| && |m_{\nu_\mu\nu_\tau}|e^{i\psi} \\
|m_{\nu_e\nu_\tau}| && |m_{\nu_\mu\nu_\tau}|e^{-i\psi} && |m_{\nu_\tau\nu_\tau}| 
\end{pmatrix*}; 
\end{equation}
with $\psi=\phi_{\mu\tau}+\phi_2=\phi_{\mu\tau}+\phi_{e\mu}-\phi_{e\tau}$.\\

{\bf Case B}: $\phi_1=\phi_{e\mu}$ and 
$\phi_2=-\phi_{\mu\tau}$. Producing
\begin{equation}\label{mngepB}
M_\nu^\prime= \begin{pmatrix*}[l] |m_{\nu_e\nu_e}| && |m_{\nu_e\nu_\mu}| && |m_{\nu_e\nu_\tau}|e^{-i\psi} \\ 
|m_{\nu_e\nu_\mu}| && |m_{\nu_\mu\nu_\mu}| && |m_{\nu_\mu\nu_\tau}| \\ 
|m_{\nu_e\nu_\tau}|e^{i\psi} && |m_{\nu_\mu\nu_\tau}| && |m_{\nu_\tau\nu_\tau}| 
\end{pmatrix*}.
\end{equation}\\

{\bf Case C}: $\phi_2=-\phi_{\mu\tau}$ and  
$\phi_1=\phi_2+\phi_{e\tau}=\phi_{e\tau}-\phi_{\mu\tau}$. Producing
\begin{equation}\label{mngepC}
M_\nu^\prime=\begin{pmatrix*}[l] |m_{\nu_e\nu_e}| && |m_{\nu_e\nu_\mu}|e^{i\psi} && |m_{\nu_e\nu_\tau}| \\ 
|m_{\nu_e\nu_\mu}|e^{-i\psi} && |m_{\nu_\mu\nu_\mu}| && |m_{\nu_\mu\nu_\tau}| \\ 
|m_{\nu_e\nu_\tau}| && |m_{\nu_\mu\nu_\tau}| && |m_{\nu_\tau\nu_\tau}| 
\end{pmatrix*}; 
\end{equation}

From the former, we can conclude that using a ``weak basis transformation.'' 
we can get rid of two unwanted phases, ending up with a single phase responsible 
for the possible CP violation phenomena present in the $U_{PMNS}$  mixing matrix.

Counting parameters once more, we find that in matrix (\ref{mngepA}) (or 
equivalently in (\ref{mngepB}) or (\ref{mngepC})), the final number 
of parameters are 
six real numbers and one phase $(\psi)$, just enough to accommodate the three mixing 
angles $\theta_{12},\theta_{13}$ and $\theta_{23}$ and the three neutrino 
masses $m_1, m_2$ and $m_3$, together with just one CP violating phase 
to take into account the CP violation in the $U_{PMNS}$ matrix via the 
parameter $(\delta_{CP}$) for Dirac neutrinos. Further texture zeros will give relationships between neutrino masses and mixing parameters.
  
In this way, one texture zero would allow us to write one of the mixing angles 
$\theta_{ij}$ as a function of the neutrino masses; meanwhile, two texture zeros allow us to 
write two mixing angles as a function of the three neutrino masses. Three or more texture zeros are meaningless.

The most important consequence of the former analysis is that since the 
phases $\phi_1$ and $\phi_2$ are arbitrary, and they can take any value, and in consequence, 
the final phase $\psi$ can be placed in any entry of the neutrino mass 
matrix, according to equations 
(\ref{mngepA})-(\ref{mngepC}). In particular, if we impose 
an off-diagonal vanishing element, we can place the phase 
$\psi$ in that entry, meaning a phase will not be present in the mass matrix. That implies CP conservation  
for that case; \rm{a result obtained from the Jarlskog invariant analysis as shown in the main text.}

\section{Appendix B}\label{appendixB}
As mentioned in the main text, the mass matrix $M_\nu$ for Dirac neutrinos, in the context of the SM enlarged with right-handed neutrinos, can be taken to be hermitian without loss of generality, which means that three independent off-diagonal matrix elements are in general complex, while three independent diagonal ones are real. If $n$ of them are taken to vanish ($n$ independent texture zeros), then a combinatorial analysis allows us to write the number of independent matrices as~\cite{liu}: 
\begin{equation}\label{ntz}
 C_n=\frac{6!}{n!(6-n)!},
\end{equation}
which means: $C_1=6$, $C_2=15$, and $C_3=20$.
(Textures with $n\geq 3$, are not realistic. )

\subsection{One texture zero}
There are two different situations: texture zero in the main diagonal and texture zero off the main diagonal, with three different cases for each situation:

\subsubsection{Diagonal texture zero}
There are three different cases given by the three matrices

\[O_1=\left(\begin{array}{ccc}0&a&b\\a^*&x_2&c\\b^*&c^*&x_3\end{array}\right),\;\;
O_2=\left(\begin{array}{ccc}x_1&a&b\\a^*&0&c\\b^*&c^*&x_3\end{array}\right),\;\;
O_3=\left(\begin{array}{ccc}x_1&a&b\\a^*&x_2&c\\b^*&c^*&0\end{array}\right),\]
all three cases related to CP violation.
\subsubsection{Off diagonal texture zero}
Again there are three different cases given by the matrices:
\[O_4=\left(\begin{array}{ccc}x_1&0&b\\0&x_2&c\\b^*&c^*&x_3\end{array}\right);\;\;
O_5=\left(\begin{array}{ccc}x_1&a&0\\a^*&x_2&c\\0&c^*&x_3\end{array}\right);\;\;
O_6=\left(\begin{array}{ccc}x_1&a&b\\a^*&x_2&0\\b^*&0&x_3\end{array}\right),\]
all of them related to CP conservation.

\subsection{Two Texture Zeros}
There are 15 different cases grouped in three different categories:


\subsubsection{One diagonal and other off-diagonal Texture zeros}
There are nine different cases 

\begin{equation}\label{eq:T11}
A_1=\begin{pmatrix*}[l]
0&0&b\\0&x_2&c\\b^*&c^*&x_3
\end{pmatrix*},
\quad
A_2=\begin{pmatrix*}[l]
0&a&0\\a^*&x_2&c\\0&c^*&x_3
\end{pmatrix*}; 
\quad 
A_3=\begin{pmatrix*}[l]
0&a&b\\a^*&x_2&0\\b^*&0&x_3
\end{pmatrix*}, 
\end{equation} 

\begin{equation}\label{eq:T22}
A_4=\begin{pmatrix*}[l]
x_1&0&b\\0&0&c\\b^*&c^*&x_3
\end{pmatrix*},
\quad
A_5=\begin{pmatrix*}[l]
x_1&a&0\\a^*&0&c\\0&c^*&x_3
\end{pmatrix*}; 
\quad 
A_6=\begin{pmatrix*}[l]
x_1&a&b\\a^*&0&0\\b^*&0&x_3
\end{pmatrix*}, 
\end{equation} 

\begin{equation}\label{eq:T33}
A_7=\begin{pmatrix*}[l]
x_1&0&b\\0&x_2&c\\b^*&c^*&0
\end{pmatrix*},
\quad
A_8=\begin{pmatrix*}[l]
x_1&a&0\\a^*&x_2&c\\0&c^*&0
\end{pmatrix*}; 
\quad 
A_9=\begin{pmatrix*}[l]
x_1&a&b\\a^*&x_2&0\\b^*&0&0
\end{pmatrix*}, 
\end{equation} 

all of them CP conserving.\\
\subsubsection{Two Texture zeros off the main diagonal}\label{dosceros}
There are three different cases

\[A_{10}=\left(\begin{array}{ccc}x_1&a&0\\a^*&x_2&0\\0&0&x_3\end{array}\right),\;\;
A_{11}=\left(\begin{array}{ccc}x_1&0&b\\0&x_2&0\\b^*&0&x_3\end{array}\right),\;\;
A_{12}=\left(\begin{array}{ccc}x_1&0&0\\0&x_2&c\\0&c^*&x_3\end{array}\right).\]

All of them are related to CP conservation. 

\subsubsection{Two Texture zeros in  the main diagonal}

There are three different cases too,

\[A_{13}=\left(\begin{array}{ccc}0&a&b\\a^*&0&c\\b^*&c^*&x_3\end{array}\right),\;\;
A_{14}=\left(\begin{array}{ccc}0&a&b\\a^*&x_2&c\\b^*&c^*&0\end{array}\right),\;\;
A_{15}=\left(\begin{array}{ccc}x_1&a&b\\a^*&0&c\\b^*&c^*&0\end{array}\right).\]
All of them related to CP violation.
\section{Appendix C}\label{appendixC}
In this appendix we review the definition of the Jarlskog's Invariant. 

The Swedish physicist Cecilia Jarlskog found that the area of each of the six unitary triangles found in a unitary matrix $3 \times 3$ (which is the same for all of them), is given by the relation: 
\[-Ar=J/2\]
with $-J$ known as Jarlskog's invariant \cite{PhysRevLett.55.1039}; which, in the parametrization that makes use of the Euler angles takes the form:  

\begin{equation}\label{jarlk}
-J= c_{12}c_{23}c_{13}^2s_{12}s_{23}s_{13}\sin\delta_{13}
\end{equation}
which can also be written as:
\begin{equation}\label{jarlkv}
 -|J|={\rm Im}(U_{ij}U_{kl}U_{kj}U_{il})
\end{equation}
for any combination of $i,j,k,l$ with $i\neq k$ y $j\neq l$.

Undoubtedly, it is the Jarlskog invariant that is the important information carrier for CP-violation \cite{Esteban:2020cvm}.

\section{Appendix D}\label{appendixD}
In this appendix we present the summary of the numerical results obtained from our analysis for the nine cases of two texture zeroes, one in the main  diagonal and the other one outside of it. For the analysis, we use $\delta_{CP}=0$. From the results, normal ordering is suggested. Three tables are presented: 

The first one correspond to the analysis done for $m_1>0$ and $m_2< 0$.  The second is for $m_1<0$ and  $m_2>0$. And in the third table are the results obtained when we relax the constraints imposed by the value $\theta_{13}$; that is, without having a correlation on the $\sin\theta_{13}$ mixing angle with the other parameters. 

\begin{table}[h!]
\begin{center}
\begin{tabular}{| c | c | c | c |c | c | c |}\hline
Texture           & $\sin^2{\theta_{12}}$ & $\sin^2{\theta_{23}}$ & $\sin^2{\theta_{13}}$ & $|m_1|$ (eV) & $|m_2|$ (eV) & $|m_3|$ (eV)\\ \hline
 1: $A_1$ & 0.298  & 0.250 & 0.022 &0.0018 &0.0091 &0.0512 \\ \hline 
 2: $A_2$ & 0.305  & 0.018 & 0.014 &0.0044 & 0.0096&0.0503 \\ \hline 
 3: $A_3$ & 0.334  & 0.007 & 0.022 & 0.0021& 0.0087& 0.0501 \\ \hline 
 4: $A_4$ & 0.318  & 0.200 & 0.022 & 0.0046 & 0.0097& 0.0512\\ \hline 
 5: $A_5$ & 0.465  & 0.010 & 0.032 & 0.0153& 0.0175&0.0534 \\ \hline 
 6: $A_6$ & 0.524  & 0.003 & 0.014 & 0.0209& 0.0227& 0.0546\\ \hline 
7: $A_7$ & 0.315  & 0.646 & 0.022 & 0.0333 &0.0344 &0.0608 \\ \hline 
 8: $A_8$ & 0.022  & 0.515 & 0.023 & 0.2685 &0.2686 &0.2731 \\ \hline 
 9: $A_9$ & 0.023  & 0.516 & 0.023& 0.2765& 0.2767  & 0.2810  \\ \hline 
\end{tabular}
\caption{Result of the mixing angles and the neutrino masses for the nine different textures obtained according to our analysis for the case $m_1>0$ and $m_2<0$.}
\label{tab:TZangles}
\end{center}
\end{table}

From table \ref{tab:TZangles} we can see that only the texture $A_7$ is able to accommodate the measured mixing  angles, with the corresponding predictions for the neutrinos mass values.

\begin{table}[h!]
\begin{center}
\begin{tabular}{| c | c | c | c |c | c | c |}\hline
Texture           & $\sin^2{\theta_{12}}$ & $\sin^2{\theta_{23}}$ & $\sin^2{\theta_{13}}$ & $|m_1|$ (eV) & $|m_2|$ (eV) & $|m_3|$ (eV)\\ \hline
 1: $A_1$ & 0.255  & 0.104 & 0.022 &0.0026 &0.0093 &0.0599 \\ \hline 
 2: $A_2$ & 0.274  & 0.102 & 0.023 &0.0040 & 0.0096&0.0513 \\ \hline 
 3: $A_3$ & 0.379  & 0.548 & 0.022 & 0.0411& 0.0420& 0.0649 \\ \hline 
 4: $A_4$ & 0.320  & 0.190 & 0.022 & 0.0041 & 0.0099& 0.0508\\ \hline 
 5: $A_5$ & 0.296  & 0.325 & 0.020 & 0.0442& 0.0450&0.0665 \\ \hline 
 6: $A_6$ & 0.572  & 0.426 & 0.021 & 0.0315 & 0.0328 & 0.0600\\ \hline 
7: $A_7$ & 0.002  & 0.542 & 0.021 & 0.0769 &0.0774 &0.0920 \\ \hline 
8: $A_8$ & 0.999  & 0.998 & 0.517 & 0.0237 &0.0253 &0.0557 \\ \hline 
 9: $A_9$ & 0.010  & 0.986 & 0.511 & 0.0241 & 0.0256  & 0.0553  \\ \hline 
\end{tabular}
\caption{Result of the mixing angles and the neutrino masses for the nine different textures obtained according to our analysis for the case $m_1<0$ and $m_2>0$.}
\label{tab:TZanglesM1}
\end{center}
\end{table}

From table \ref{tab:TZanglesM1} we can see that only the texture $A_3$ is able to accommodate the measured mixing  angles, with the corresponding predictions for the neutrinos mass values.


To compare our results with previous published studies, we have carried through one alternative analysis 
consisting in relaxing the constraint imposed by the value $\theta_{13}$; that is,  without having a correlation of the $\sin{\theta_{13}}$ mixing angle with the other parameters. The results obtained are presented in Table \ref{tab:Tpeinado} of this appendix. From these values we can see that textures $A_1$, $A_4$, $A_7$ and $A_8$ are in fairly good agreement with the experimental measured numbers at $3\sigma$, results in agreement with the analysis already presented in Refs.~\cite{lenis2023twozero, liu}. 
It is important  to notice that when the parameter $\sin{\theta_{13}}$ is smoothed, it is not possible to make predictions about all the entries in the $U_{PMNS}$ mixing matrix.
\begin{table}[h!]
\begin{center}
\begin{tabular}{| c | c | c | c | c | c |}\hline
Texture           & $\sin^2{\theta_{12}}$ & $\sin^2{\theta_{23}}$  & $|m_1|$ (eV) & $|m_2|$ (eV) & $|m_3|$ (eV)\\ \hline
 1: $A_1$ & 0.320  & 0.522 & 0.0077 &0.0114 &0.0514 \\ \hline 
 2: $A_2$ & 0.215  & 0.318 & 0.0089 & 0.0126&0.0508 \\ \hline 
 3: $A_3$ & 0.334   & 0.0146 & 0 & 0.0089  & 0.0497  \\ \hline 
 4: $A_4$ & 0.320  & 0.522 & 0.0355& 0.0365& 0.0611 \\ \hline 
 5: $A_5$ & 0.230  & 0.430 & 0 & 0.0084& 0.0512 \\ \hline 
 6: $A_6$ & 0.540  & 0.020 & 0.0230 & 0.0245 & 0.0551 \\ \hline 
 7: $A_7$ & 0.325  & 0.657 & 0.0283 & 0.0297& 0.0569\\ \hline 
 8: $A_8$ & 0.318  & 0.512 & 0 & 0.0087 & 0.0511\\ \hline  
  9: $A_9$ & 0.364  & 0.909& 0.0303 & 0.0585  & 0.0585  \\ \hline 
\end{tabular}
\caption{Values of the parameters obtained for the nine different two zero textures for the alternative analysis.}\label{tab:Tpeinado}
\end{center}
\end{table}


\pagebreak

\bibliographystyle{unsrt}
\bibliography{bilbliografia}

\end{document}